\documentclass[prb,twocolumn,showpacs,preprintnumbers,amsmath,amssymb,floatfix]{revtex4-1}
\usepackage{graphicx}
\usepackage{dcolumn}
\usepackage{pdfpages}
\usepackage{hyperref}
\usepackage{bm}
\usepackage{longtable}
\usepackage[a4paper, margin=0.7in]{geometry}
\begin{document}
\author{Hirak Kumar Chandra$^{1}$}
\email{Email: hirakkumar02@gmail.com}
\author{Guang-Yu Guo$^1$}
\affiliation{$^1$Department of Physics, National Taiwan University, No. 1, Sec. 4, Roosevelt Rd., Taipei 10617, Taiwan (R.O.C.)}
\title {Topological insulator associated with quantum anomalous Hall phase in ferromagnetic perovskite superlattices} 

\begin{abstract}
We do a search for topological insulators which are associated with ferromagnetic ordering and show anomalous quantum Hall effect, among transition metal oxide superlattices taking the parent compounds as LaAlO$_3$ and SrTiO$_3$. Among the various superlattices which are studied here, (LaAlO$_3$)$_{10}$/(LaOsO$_3$)$_2$ exhibits a ferromagnetic ground state with a topologically non-trivial energy gap when a spin-orbit interaction is turned on. The study of transverse conductivity shows that the system has quantized Hall conductivity inside the topological energy gap without  applying any external magnetic field. The ferromagnetic order parameters and the ordering temperature ($T_c$) have been estimated by taking a simple Heisenberg model of ferromagnetism.     
\end{abstract}

\maketitle
\section{Introduction}
A quantum state of matter, which has a bulk energy band gap separating the valence band maximum (VBM) and the conduction band minimum (CBM) like an insulator in its interior but behaves like a metal on its boundary, is classified as topological insulator (TI). The gapless states of a TI on its boundary (surface) are protected by time-reversal (TR) symmetry. The surface states of a TI give rise to conducting states, which are predicted to have special properties useful for spintronics and quantum computations. The topological insulating property of a material is closely associated with integer quantum Hall effect (QHE). QHE arises in a two-dimensional (2D) electron gas, in the presence of an external magnetic field, where the energy spectrum of electron splits into Landau levels. Haldane, in his novel work, showed that QHE may also arise even in the absence of a net magnetic flux within a unitcell when the TR symmetry is broken in a periodic 2D system\cite{Haldane}. It was shown that, the ground state of a periodic 2D electron system having a gap in the single particle density of states at Fermi level possesses quantized transverse conductivity ($\sigma_{xy}$) $\frac{ne^2}{h}$, where $n$ is generally rational but can be integer in the absence of electron-electron  interactions\cite{Thouless}.

	In Haldane's model, the spin degrees of freedom of the electrons was not considered and letter on it was generalized to an electron system having spins. In this model the electronic states become doubly degenerate when there is no coupling between up-spin and down-spin states. The electrons in the conducting edge channel are chiral, i.e., flowing counter clockwise when $n=1$ and clockwise when $n=-1$ which is the characteristic of the topology of bulk quantum Hall state. Kane and Mele generalized the Haldane model including electron spin degrees of freedom for a graphene lattice\cite{Kane}. They introduced spin-orbit interactions between electron spin and momentum and showed that graphene converts from a 2D semi-metallic state to a quantum spin Hall (QSH) insulator. Here, unlike the QHE where the TR symmetry is broken by periodic magnetic filed, the spin-orbit coupling (SOC) preserves TR symmetry. The edge states are nonchiral in the case of QSH insulator and the up-spin electrons in the edge channel flow in one direction while the down spin-electrons flow in the opposite direction\cite{Wu,Xu}. As a result, the Hall conductivity becomes zero but there is a quantized spin Hall conductivity ($\sigma^{s}_{xy}$), defined by $J^{\uparrow}_{x}-J^{\downarrow}_{x}=\sigma^{s}_{xy}E_{y}$ with $\sigma^{s}_{xy}=\frac{e}{2\pi}$, where $J^{\uparrow}_{x}$ and $J^{\downarrow}_{x}$ are up-spin and down-spin currents respectively along $x$ direction and $E_{y}$ is the electric filed along $y$ direction. Kane and Mele also showed that under the TR symmetry the edge states in the QSH insulator is robust against the violation of spin conservation because of their crossing at Brillouin zone boundary leads to Kramers doublet which remains invariant for any TR symmetric perturbation. This proves that the QSH insulator is a topological phase of matter.
	
	Graphene is made of carbon atoms which is a light element and said to have a weak SOC strength. In the search of strong SOC strengths made from heavy elements Bernevig $et$ $al.$\cite{BHZ} proposed a quantum well model of HgTe/CdTe. They showed that the transition from a conventional insulating phase to a topological quantum phase occurs for a critical well thickness ($d_c$). Their proposal leads to the experimental detection of QSH insulator phase. They considered a quantum well structure where HgTe was sandwiched between layers of CdTe and with the $k.p$ theory they established that band inversion takes place at the $\Gamma$ point in the Brillouin zone above a particular thickness ($d_c=6.3 nm$). This band inversion leads to a quantum phase transition from a trivial insulator to a QSH insulator. Soon after the theoretical prediction of QSH in HgTe/CdTe quantum well, K\"{o}nig $et$ $al.$\cite{konig} grew thin HgTe quantum well sandwiched between (Hg,Cd)Te barriers using molecular beam epitaxy to observe the phenomena of QSH effect experimentally. They observed a normal electronic structure having zero conductivity for samples with narrow quantum well whereas an inverted electronic structure with a conductivity close to the conductivity of a QSH insulator was found above a critical well thickness($d_c$). 
	
	We discussed earlier that the topological surface stares are protected by TR symmetry. The Dirac nodes on topological surface states are robust in the presence of non-magnetic disorder because this perturbation can not break the TR symmetry. But, the TR symmetry can be broken by introducing a magnetic impurity, what is known as quantum anomalous Hall (QAH) effect. QAH effect and the associated disorder, which may lift the degeneracy at the Dirac points, opens up a band gap in the surface states. This effect was theoretically predicted in Mn doped HgTe quantum well where the quantized Hall conductivity was obtained for a range of well thickness and the doping concentration\cite{cxliu}. Another theoretical work based on first-principle calculations predicted that thin films of tetradymite semiconductors (e.g. Bi$_2$Te$_3$, Bi$_2$Se$_3$, Sb$_2$Te$_3$) doped with transition metal elements (Cr or Fe) would be possible candidates to show QAH state\cite{rui_wu}. An experimental observation of QAH in magnetic TI was reported by Chang $et$ $al.$ in the case of Cr doped (BiSb)$_2$Te$_3$ thin films\cite{chang}. In another observation of QAH, it was shown that Fe on the surface of Bi$_2$O$_3$ opens up a gap at the Dirac point, whose value is likely to be set by the interaction of Fe ions with the Se surface, due to the disorder induced TR symmetry breaking on the surface\cite{wary}.                                            

	In the search of TI materials, transition metal oxides (TMOs) were also predicted to be possible candidates with a non-trivial band topology. TMOs show a variety of interesting properties depending on their electron-electron correlations such as magnetism, ferroelectricity, superconductivity, Mott insulators etc. Based on first-principle and tight-binding (TB) calculations, it was proposed that bilayers of perovskite TMOs grown along [111] direction are potential candidates for 2D TIs\cite{Xiao1}. The topological band structures and the associated QSH phase of these materials were predicted to be depended on substrates, dopant ions and external gate voltages\cite{Xiao2}. Recent theoretical works suggested that anti-perovskite TMOs (A$_3$BO) may also become promising candidates for new TIs via their interplay between band topology, crystal symmetry and electron-electron correlations\cite{Timothy}. TI phase associated with QAH state could also be realized in double-perovskite (La$_2$MnIrO$_6$) monolayers grown along [001] direction as theoretically suggested by Zhang $et$ $al.$ via their first-principle and TB calculations\cite{Hongbin}. In our theoretical work based on first-principle calculations we do a search for TI phase associated with a magnetic ground state among perovskite TMO bilayers [111]. Here, we investigate which of the TMO bilayers (superlattices) have non-trivial topological band gap associated with QAH phase while SOC is turned on. To establish the topologically non-trivial nature of band gap, anomalous Hall conductivity ($\sigma^{AH}_{xy}$) has been estimated as a function of energy near the Fermi level. Applying Heisenberg model of ferromagnetism, we have estimated the inter-atomic exchange parameters as well as magnetic ordering temperature to establish the magnetic ground state.       

\section{Methodology}
Cubic perovskite oxides ABO$_3$ (LaAlO$_3$, SrTiO$_3$) have been taken here as the parent compounds. The experimental lattice constant of LaAlO$_3$ and SrTiO$_3$ have been optimized with the first-principle calculations based on density functional theory (DFT) as implemented in VASP\cite{kresse1}. The optimized lattice constant ($a_0$) for LaAlO$_3$ comes out to be 3.81 \AA{} which is close to its experimental value, 3.79 \AA\cite{Geller} as well as for the case of SrTiO$_3$ also, the optimized lattice constant (3.94 \AA) is observed to be close to its experimental value (3.91 \AA\cite{Megaw}).

\begin{figure}[h]
\centerline{\includegraphics[angle=0,scale=0.35,clip]{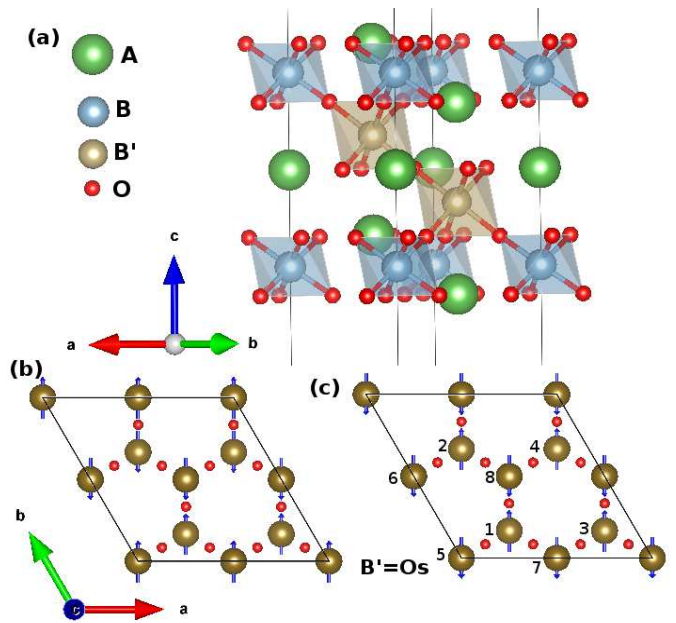}}
\caption{(color online) (a) A truncated picture of ABO$_3$ superlattice grown along [111], where two of its 12 layers have been substituted by AB$^\prime$O$_3$. Each type of atom has been indicated with different colour. Two types of AFM configurations viz. (b) AFM1 and (c) AFM2, within Os networks of (LaAlO$_3$)$_{10}$/(LaOsO$_3$)$_2$ superlattice, are shown. Up and down arrows indicate the up and down spin moments assign to the respective Os atoms respectively. Atoms indicated by 1, 2, 3 and 4 lie in one plain whereas 5, 6, 7 and 8 lie in another plain along $c$ direction.} 
\label{Fig1}
\end{figure}	

	A sixty-atom supercell of ABO$_3$, grown along [111] direction, has been constructed which consists of 12 AO$_3$ layers and 12 B layers. A cubic lattice forms a honeycomb lattice when projected along the (111) plain with lattice parameters, $a_{in-plain}=\sqrt{2}a_0$ and $a_{out-of-plain}=\sqrt{3}a_0$, where $a_0$ is the cubic lattice constant. Two consecutive B layers have been substituted by other transition metal elements (B$^\prime$= Ag,Au,Os,Re,Ir or Rh) to construct a superlattice consisting of 10 layers of ABO$_3$ and 2 layers of AB$^\prime$O$_3$ (Fig.\ref{Fig1}). To get the minimum energy configurations, the internal coordinates of each superlattice has been been relaxed  till the force on individual atom becomes $\leq$10 meV/\AA. First-principle density functional calculations, as implemented in VASP, have been carried out to solve the electronic structure of each of the present systems. A k-point mesh of $12\times 12\times 2$ with $\Gamma$ symmetry has been taken for the calculations and a plain wave cutoff energy of 400 eV has been used. Pseudopotentials based on the projector augmented wave (PAW)\cite{kresse2} with GGA-PBE\cite{perdew} exchange have been taken for our calculations. Band structures of each bilayer has been plotted without and with spin-orbit coupling (SOC) implementation\cite{kresse3}, setting the magnetization axis along $c$ ([111]) direction. Calculations including SOC have been performed in the noncollinear mode as implemented in VASP by Hobbs $et$ $al.$\cite{Hobbs} and Marsman and Hafner\cite{Marsman}. Taking the case of (LaAlO$_3$)$_{10}$/(LaOsO$_3$)$_2$, which shows a topological band gap\cite{Xiao1} with a finite magnetic moment, the anomalous Hall conductivity (AHC) has been calculated using Berry phase formalism. Based on this formalism, AHC ($\sigma^{AH}_{xy}$) is given by a sum over the Berry curvatures on all k-points within the Brillouin zone for all the occupied bands,
\begin{equation}
\sigma^{AH}_{xy} = \frac{e^2}{h}\frac{1}{N_{k}\Omega_{c}}\sum_{k}(-1)\Omega_{xy}(\textbf{k}), 
\label{ahc}
\end{equation}
\begin{equation}
\Omega_{xy} = \sum_{n}f_{n\textbf{k}}\Omega_{n,xy}(\textbf{k}), 
\label{berry}
\end{equation}
where, $N_k$ and $\Omega_c$ are the number of $k$-points and volume of the cell respectively, and $f_{n\textbf{k}}$ and $\Omega_{n,xy}$ are the Fermi distribution function and the Berry curvature for the $n$th band at $\textbf{k}$ respectively. The berry curvature can be written in terms of an antisymmetric tensor as,
\begin{equation}
\Omega_{n,xy}(\textbf{k}) = \epsilon_{xyz}\Omega_{nz}(\textbf{k}) = -2Im\langle\nabla_{k_{x}}u_{n\textbf{k}}\vert\nabla_{k_{y}}u_{n\textbf{k}}\rangle ,
\end{equation}  
where, $\langle u_{n\textbf{k}}\rangle =e^{-i\textbf{k}.\textbf{r}}\vert\Psi_{n\textbf{k}}\rangle$ represents the cell-periodic Bloch states. By setting the Wannier functions\cite{Marzari} spanning the occupied bands (together with a few low lying unoccupied bands, typically), the AHC, as given in Eq.(\ref{ahc}), can be evaluated via Wannier interpolation\cite{x_wang,lopez} without any truncation. 

The inter-atomic exchange coupling parameters as well as the magnetic ordering temperature ($T_C$) for the system have been estimated from the ground state energies of ferromagnetic and anti-ferromagnetic spin configurations using the Heisenberg model. Two types of anti-ferromagnetic configurations (AFM1 and AFM2) have been considered which are shown in fig.1(b) and (c). The magnetic anisotropy energy (MAE) has also been calculated taking the magnetic axis once along $c$ ([111]) direction and then $a$ ([110]) direction.   

\section{Results and discussion}
	In the search of TI phase associated with QAH effect in perovskite TMO bilayers (superlattices) which have magnetic ground states, we start with a band structure that possesses Dirac points in Brillouin zone without SOC. We have taken similar TMO bilayers as mentioned in Ref.\cite{Xiao1} and examined which of those have magnetic ground states. Form our first principle, spin-polarized DFT calculations, it has been found that, (LaAlO$_3$)$_{10}$/(LaReO$_3$)$_2$, (LaAlO$_3$)$_{10}$/(LaOsO$_3$)$_2$, (SrTiO$_3$)$_{10}$/(SrRhO$_3$)$_2$, (SrTiO$_3$)$_{10}$/(SrOsO$_3$)$_2$ and (SrTiO$_3$)$_{10}$/(SrIrO$_3$)$_2$ superlattices have magnetic ground states. Next we examine whether the band gap could be opened up at those Dirac points with turning on the SOC interactions. The physical properties of the superlattices, as mentioned here,tabulated in the following table (Table.\ref{MI}) which have been obtained from the non-collinear DFT calculations with SOC. 

\begin{table}[h]\caption{Physical properties of the superlattices taken in the present work. LAO and SRO represents LaAlO$_3$ and SrTiO$_3$ respectively. Zero band gap indicates the system is metallic.}
\begin{tabular}{|c|c|c|c|c|}
\hline
Superlattice  & Band & Magnetic & Magnet-   \\
			 & gap(meV) & state & ization ($\mu_B$) \\
\hline
(LAO)$_{10}$/(LaAgO$_3$)$_2$ & 24 & AFM & 0.00  \\
(LAO)$_{10}$/(LaAuO$_3$)$_2$ & 99 & AFM & 0.00  \\
(LAO)$_{10}$/(LaReO$_3$)$_2$ & 0  & FM  & 0.92  \\
(LAO)$_{10}$/(LaOsO$_3$)$_2$ & 38 & FM  & 0.83  \\
(STO)$_{10}$/(SrAgO$_3$)$_2$ & 0  & NM  & 0.00  \\
(STO)$_{10}$/(SrRhO$_3$)$_2$ & 0  & FM  & 1.89  \\
(STO)$_{10}$/(SrOsO$_3$)$_2$ & 0  & FM  & 2.66  \\ 
(STO)$_{10}$/(SrIrO$_3$)$_2$ & 0  & FM  & 0.08  \\
\hline
\end{tabular}
\label{MI}
\end{table}

\begin{figure}[h]
\centerline{\includegraphics[angle=0,scale=0.36,clip]{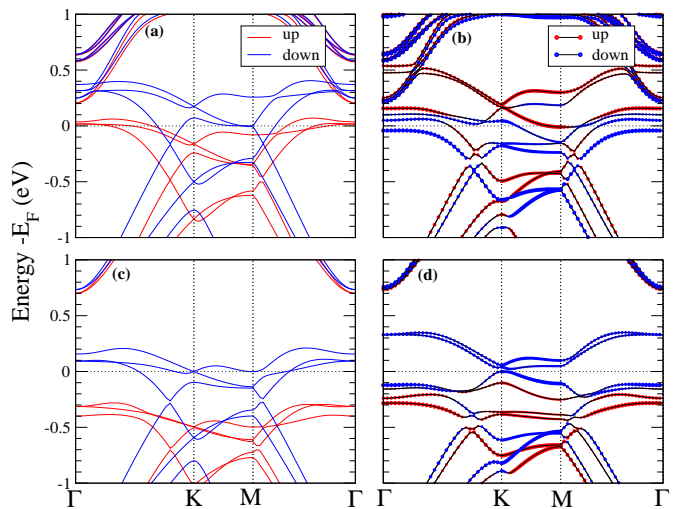}}
\caption{(color online) The band dispersions are plotted for (a),(b) (LaAlO$_3$)$_{10}$/(LaReO$_3$)$_2$  and (c),(d) (LaAlO$_3$)$_{10}$/(LaOsO$_3$)$_2$ superlattices respectively. The left and the right column show band dispersion plots without and with SOC respectively. The energy axis for each case is set by taking the Fermi energy as reference level.} 
\label{Fig2}
\end{figure}

From Table \ref{MI}, one can find that only (LaAlO$_3$)$_{10}$/(LaOsO$_3$)$_2$ superlattice, among the superlattices which are mentioned here, is coming out to have insulating and ferromagnetic (FM) ground state. The magnetization of each of the superlattices are also tabulated (Table.\ref{MI}). The band dispersions without SOC (but with spin-polarized) and with SOC have been plotted for each of the superlattices to establish the above mentioned physical properties. First of all, the band dispersion plot of (LaAlO$_3$)$_{10}$/(LaReO$_3$)$_2$ and (LaAlO$_3$)$_{10}$/(LaOsO$_3$)$_2$ without and with SOC  are shown in fig.\ref{Fig2}. The energy axis has been set by taking Fermi energy as the reference level. In fig.\ref{Fig2}, panel (a) and (b) show the band dispersion plot of  (LaAlO$_3$)$_{10}$/(LaReO$_3$)$_2$ without and with SOC respectively whereas panel (c) and (d) show the band dispersion plots of (LaAlO$_3$)$_{10}$/(LaOsO$_3$)$_2$ without and with SOC respectively. The up and down spin bands are indicated by red and blue colours respectively. From this figure it can be seen that without SOC (LaAlO$_3$)$_{10}$/(LaReO$_3$)$_2$ has a metallic ground state whereas (LaAlO$_3$)$_{10}$/(LaOsO$_3$)$_2$ has a half-metallic ground state with a Dirac point near "K" in the Brillouin zone. With the application of SOC no band gap opens up in the case of (LaAlO$_3$)$_{10}$/(LaReO$_3$)$_2$ but a gap of 38 meV opens up near "K" in the case of (LaAlO$_3$)$_{10}$/(LaOsO$_3$)$_2$. The magnetization in this case with SOC interaction is found to be 0.83 $\mu_B$ along $c$ direction. 

\begin{figure}[h]
\centerline{\includegraphics[angle=0,scale=0.36,clip]{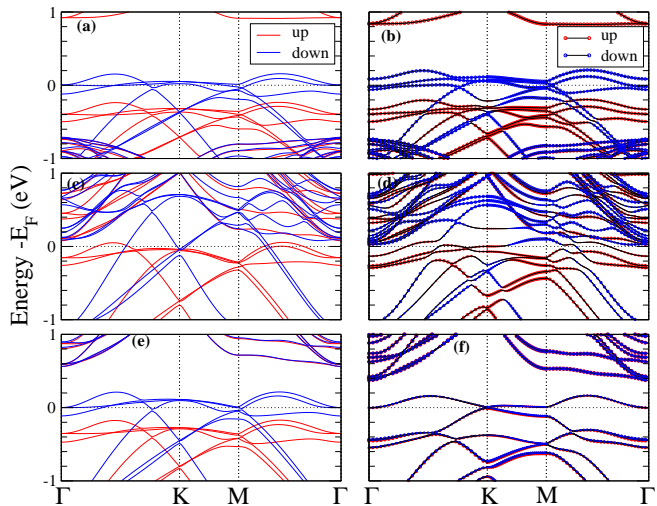}}
\caption{(color online) The band dispersions are plotted for (a),(b) (SrTiO$_3$)$_{10}$/(SrRhO$_3$)$_2$ , (c),(d) (SrTiO$_3$)$_{10}$/(SrOsO$_3$)$_2$ and (e),(f) (SrTiO$_3$)$_{10}$/(SrIrO$_3$)$_2$ superlattices respectively. The left and the right column show band dispersion plots without and with SOC respectively. The energy axis for each case is set by taking Fermi energy as reference level. Up and down spin bands are indicated by red and blue colours respectively.}
\label{Fig3}
\end{figure}

	Next, based on SrTiO$_3$ compound, the band dispersions, without and with SOC, of three superlattices which are mentioned in Table \ref{MI} have been plotted. These plots are shown in fig.\ref{Fig3}. In fig.\ref{Fig3}, panel (a) and (b) show the band dispersion plots of (SrTiO$_3$)$_{10}$/(SrRhO$_3$)$_2$ without and with SOC respectively. Here, although (SrTiO$_3$)$_{10}$/(SrRhO$_3$)$_2$  has a half metallic ferromagnetic ground state within the spin-polarized calculations, SOC cannot open a band gap unlike (LaAlO$_3$)$_{10}$/(LaOsO$_3$)$_2$ superlattice. A magnetic moment of 1.89 $\mu_B$ has been found in this case with SOC implementation. In the same figure, panel (c) and (d) show the band dispersion plots of (SrTiO$_3$)$_{10}$/(SrOsO$_3$)$_2$ superlattice without and with SOC respectively. Here the system has a fully metallic and ferromagnetic ground state within spin-polarized calculations which remains unaltered with SOC implementation. In a similar manner panel (e) and  (f) show the band dispersion plot of (SrTiO$_3$)$_{10}$/(SrIrO$_3$)$_2$ without and with SOC respectively. Here also, one can observe the half metallic and ferromagnetic ground state within spin-polarized calculations whereas, with SOC implementation it becomes fully metallic having almost spin-degenerate band dispersion near the Fermi energy. 
	
\begin{figure}[h]
\centerline{\includegraphics[angle=0,scale=0.36,clip]{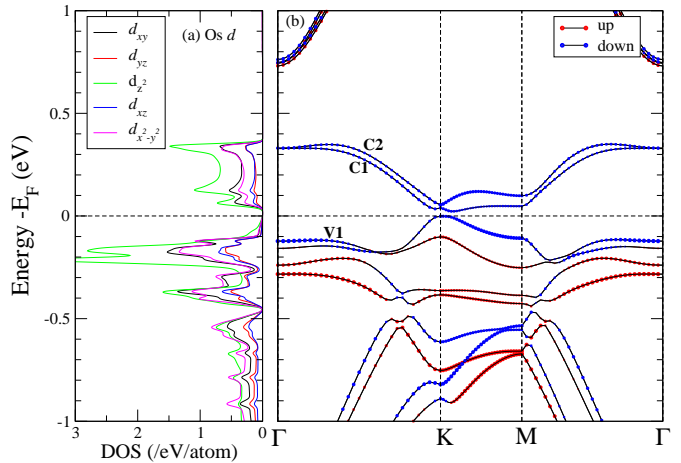}}
\caption{(color online) The Os $d$ projected orbital resolved partial density of states are plotted for (LaAlO$_3$)$_{10}$/(LaOsO$_3$)$_2$ superlattice in the left panel and spin resolved band dispersion of the same material is plotted in right panel.}
\label{Fig4}
\end{figure}

	Now we focus on (LaAlO$_3$)$_{10}$/(LaOsO$_3$)$_2$ superlattice, where a band gap of 38 meV is observed at the Dirac point as we turn on the SOC as well as it shows spin-splitted states associated with a net magnetization of 0.83$\mu_B$ along the $c$ direction. The orbital resolved Os $d$ projected partial density of states and the spin resolved band dispersion plots are shown in fig.\ref{Fig4}. The relative weight of up and down spin contribution in band dispersion are indicated by red and blue bubbles respectively and the Fermi energy is set as the reference level on the energy axis. From the band dispersion plot (Fig.\ref{Fig4}(b)), it is observed that Os $d$ up-spin states are fully filled up whereas the down spin states are splitted up near the Fermi level. The Fermi level lies within the splitted states which give rise to an energy band gap. The orbital resolved Os $d$ projected density of sates, which are shown in fig.\ref{Fig4}(a), show the contribution of five $d$ orbitals near the Fermi energy.  

To observe the relative contribution of the individual components of $d$ orbitals of Os ion, we have tabulated their relative weight for three bands near the Fermi level, as indicated in fig.\ref{Fig4}(b) (V1,C1,C2), at $\Gamma$ point.    

\begin{table}[h]\caption{Fractional contribution, in percentage, of the components of Os $d$ orbitals in the topmost valance band (V1) and two lowest conduction bands (C1 and C2) as shown in fig.\ref{Fig4} are tabulated.}
\begin{tabular}{|c|c|c|c|c|c|}
\hline
Band & $d_{xy}$ & $d_{yz}$ & $d_{xz}$ & $d_{x^2-y^2}$ & $d_{z^2}$ \\
\hline
V1	 & 23.53 & 9.46 & 9.46 & 23.53 & 0     \\
C1	 & 10.96 & 4.57 & 4.57 & 10.96 & 37.21 \\
C2	 & 18.56 & 8.48 & 8.48 & 18.56 & 12.83 \\ 
\hline
\end{tabular}
\label{band}
\end{table}
	
\begin{figure}[!h]
\centerline{\includegraphics[angle=0,scale=0.36,clip]{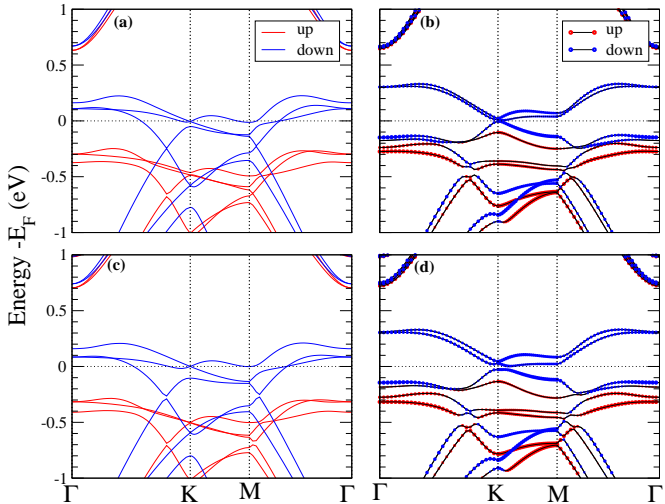}}
\caption{(color online) The band dispersions are plotted for (a),(b)(LaAlO$_3$)$_{7}$/(LaOsO$_3$)$_2$ and (c),(d) (LaAlO$_3$)$_{13}$/(LaOsO$_3$)$_2$ superlattices respectively. The left and the right column show band dispersion plots without and with SOC respectively.}
\label{Fig5}
\end{figure}

	As a function of layer thickness of the parent compound, we also examine the nature of (LaAlO$_3$)$_{n}$/(LaOsO$_3$)$_2$ superlattices, where $n$ is an integer number. We have considered two superlattices  (LaAlO$_3$)$_{7}$/(LaOsO$_3$)$_2$ and (LaAlO$_3$)$_{13}$/(LaOsO$_3$)$_2$. In the first case a reduced band gap of 10 meV has been observed associated with a reduced magnetization of 0.66 $\mu_B$. But with a thick layer of LaAlO$_3$ consisting of 13 layers, a band gap of 44 meV with a magnetization 0.85 $\mu_B$ has been observed (Fig.\ref{Fig5}). 

\begin{figure}[h]
\centerline{\includegraphics[angle=0,scale=0.35,clip]{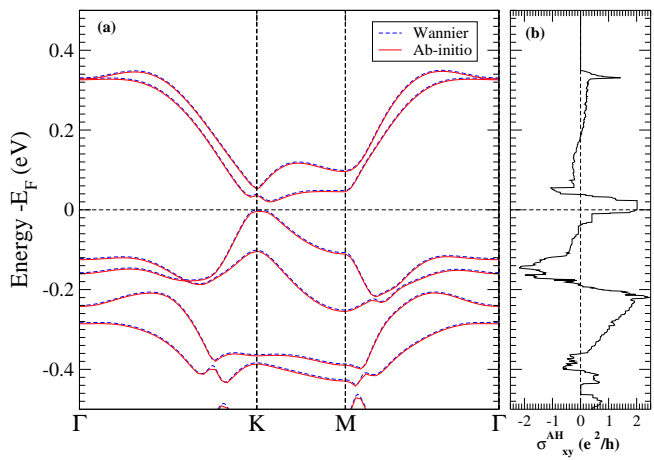}}
\caption{(color online) (a) The Wannier interpolated band structure of (LaAlO$_3$)$_{10}$/(LaOsO$_3$)$_2$ has been plotted over the $ab$-initio band structure for checking proper interpolation of Wannier functions. (b) The AHC, $\sigma^{AH}_{xy}$ (in $\frac{e^2}{h}$) has been plotted as a function of energy taking Fermi energy as the reference level.}
\label{Fig8}
\end{figure}

	To obtain the AHC ($\sigma^{AH}_{xy}$) of the present system ((LaAlO$_3$)$_{10}$/(LaOsO$_{3}$)$_{2}$), we have taken the Wannier interpolation of Os $d$ orbitals within a energy range of 2.2 eV below the Fermi energy to 1.0 eV above the Fermi energy keeping the magnetization axis along $c$ direction. The AHC ($\sigma^{AH}_{xy}$) has been evaluated within the specified energy range taking a dense k-point grid of $144\times{}144{}\times12$ within the Brillouin zone and plotted as a function of energy with setting Fermi energy at zero on the energy axis (Fig.\ref{Fig8}). The $\sigma^{AH}_{xy}$ at Fermi level has been found to be $\sim$2.01 times $\frac{e^2}{h}$, which is close to 2 (integer number) and hence we may claim that $\sigma^{AH}_{xy}(E_F)$ is quantized. 
	
	(LaAlO$_3$)$_{10}$/(LaOsO$_3$)$_2$ superlattice has been found to have a ferromagnetic ground state without SOC. Hence, we have estimated the inter-atomic exchange parameters for this superlattice based the Heisenberg model of magnetism. We have decomposed the total ground state energy ($E$) of a magnetic configuration into two parts, viz. (i) nonmagnetic energy ($E_0$) and (ii) energy due to inter-atomic exchange.                        
\begin{equation}
E = E_{0} + \sum_{i>j}J_{ij}{\bf S_{i}}.{\bf S_{j}},
\end{equation}	
	
where, $J_{ij}$ is the inter-atomic exchange parameter between sites $i$ and $j$ and {\bf S$_{i}$} and {\bf S$_{j}$} are the spins on sites $i$ and $j$ respectively. Here we have estimated the inter-atomic exchange parameters upto second neighbour. To estimate these exchange parameters we have considered three types of magnetic configurations within two layers of LaOsO$_3$. First, ferromagnetic interactions have been taken between the in-plain as well as out-of-plain Os atoms. Secondly, in an equilateral triangle formed by three in-plain Os atoms, two have been taken to have parallel spins and the third one has been taken to have anti-parallel to the first two. The immediate out-of-plain Os atom which is equidistant from the three corners of the equilateral triangle has been taken to have parallel to the first two Os atoms as mentioned above and anti-parallel to the third one (Fig.\ref{Fig1}(b)). Thirdly,  ferromagnetic interaction has been taken between in-plain Os atoms and anti-ferromagnetic interaction has been taken between out-of-plain Os atoms(Fig.\ref{Fig1}(c)). The estimated values of $J_1$ (out-of-plain exchange parameter) and $J_2$ (in-plain exchange parameter) are coming out to be -22.08 meV and -1.69 meV respectively, which indicate both the in-plain and out-of-plain Os atoms are ferromagnetically coupled. Considering only the most significant exchange parameter $J_1$, the ferromagnetic ordering temperature is estimated to be 256 K. Magnetic anisotropy energy (MAE) for (LaAlO$_3$)$_{10}$/(LaOsO$_3$)$_2$ superlattice has been calculated by taking the magnetic axis once along $c$ direction and then along $a$ direction and it comes out to be 2.5 meV considering the k-mesh convergence criterion.      

\section{Conclusion}
	In the search of topological insulator with non-trivial topological surface states associated with a magnetic ground state, we have studied bilayers of perovskite type transition metal oxides grown along [111] direction. Among the various transition metal oxide superlattices studied in this paper, only (LaAlO$_3$)$_{10}$/(LaOsO$_3$)$_2$ is coming out to have an insulating ground state with SOC, associated with a ferromagnetic ordering. The study of transverse conductivity shows that it exhibits the property of quantum anomalous Hall insulator. The inter-atomic exchange parameters have been estimated by taking the Heisenberg model of ferromagnetism. Both of the significant exchange parameters ($J_1$ and $J_2$) suggest that the system has a ferromagnetic ground state with an ordering temperature of 256 K.      

\section{Acknowledgement}
	The authors thank National Taiwan University, Taiwan for providing the necessary financial support to carry out the project.

\end{document}